\documentclass[11pt,titlepage]{amsart}
 \usepackage{amsaddr}

\usepackage{graphicx}        
 
\usepackage[table]{xcolor}
\usepackage{soul}
\usepackage{textcomp}
\usepackage{multirow} 
\usepackage{xspace}
\usepackage{amsmath} 
\usepackage{amssymb}
\usepackage{breqn}
\usepackage[bookmarks=false]{hyperref}
\usepackage{algorithm}
\usepackage{algorithmic}
\usepackage{subfigure}
\usepackage{fancyhdr}
\def\V{\mbox{$\mathcal{V}$}} %

\def\myalgo{\textsf{ML-LCD}\xspace}

\begin{document}

\title[Node-centric community detection in multilayer networks]{Node-centric community detection in multilayer networks with layer-coverage diversification bias}
\author{Roberto Interdonato}
\address{Dept. of Computer, Modeling, Electronics, and Systems Engineering, University of Calabria, Italy}  \email{rinterdonato@dimes.unical.it} 

\author{Andrea Tagarelli}
\address{Dept. of Computer, Modeling, Electronics, and Systems Engineering, University of Calabria, Italy} \email{andrea.tagarelli@unical.it}

\author{Dino Ienco}
\address{IRSTEA - UMR TETIS, France}  \email{dino.ienco@irstea.fr} 

\author{Arnaud Sallaberry}
\address{LIRMM - Universit\'e  Paul Val\'ery, France} \email{arnaud.sallaberry@lirmm.fr}

\author{Pascal Poncelet}
\address{LIRMM - Universit\'e de Montpellier, France} \email{pascal.poncelet@lirmm.fr}

\maketitle

\begin{abstract}
The problem of node-centric, or \textit{local}, community detection in information networks refers to the identification of a community for a given input node, having limited information about the network topology.  
Existing methods for solving this problem, however, are not conceived to work on complex networks. In this paper, we propose a novel framework for local community detection based on the multilayer network model. 
Our approach relies on the maximization of the ratio between the community internal connection density and the external connection density, according to  multilayer similarity-based community relations.  We also define a biasing scheme that allows the  discovery of local communities characterized by different degrees of layer-coverage diversification. Experimental evaluation  conducted  on real-world multilayer networks has shown the significance of our approach.  
\end{abstract}

\section{Introduction}\label{sec:intro}  
The classic problem of community detection in a network graph  corresponds to an optimization problem which is \textit{global} as   it requires knowledge on the \textit{whole} network structure. The problem is known to be   computationally difficult to solve, while its approximate solutions have to cope with both accuracy and efficiency issues that become more severe as the network increases in size. 
Large-scale, web-based environments have indeed traditionally represented a natural scenario for the development and testing of effective community detection approaches. 
In the last few years, the problem has attracted increasing attention in research contexts related to \textit{complex networks}~\cite{Mucha10,CarchioloLMM10,PapalexakisAI13,Kivela+14,DeDomenico15,Loe15,KimL15,Peixoto15}, 
whose modeling and analysis is widely recognized as a useful tool to better understand the characteristics and dynamics of multiple, interconnected types of node relations and interactions~\cite{BerlingerioPC13,Magnanibook}.

Nevertheless, especially in social computing, one important aspect to consider   
is that  we might often want to identify the personalized network of social contacts of interest to a single user only. To this aim, we would like  to determine the   expanded  neighborhood  of that user which  forms  a densely connected, relatively small subgraph. 
This is known as \textit{local community detection} problem~\cite{Clauset05,ChenZG09}, whose general objective is, given limited information about the network, to identify a   community  structure which is centered on one or few  seed users.   
Existing studies on this problem have focused, however, on social networks that are built on a single user relation type or context~\cite{ChenZG09,ZakrzewskaB15}. 
As a consequence, they are not able to profitably exploit the fact that most  individuals nowadays have multiple accounts across different social networks, or that relations of different types (i.e., online as well as offline relations) can be  available for the same population of a social network~\cite{Magnanibook}. 

In this work, we propose a novel framework based on the multilayer network model for the problem  of local community detection, which overcomes   the aforementioned limitations in the literature, i.e., community detection on a multilayer network but from a global perspective, and local community detection but limited to monoplex networks. 
We  have recently brought the local community detection problem into the context of multilayer networks~\cite{ASONAM16}, by providing  a preliminary formulation  based on an unsupervised approach.  A key aspect of our proposal is  the definition of similarity-based community relations that exploit  both internal and external connectivity of the nodes in the community being constructed for a given seed, while accounting for  different layer-specific topological information. 
Here we push forward our research by  introducing a parametric control in the similarity-based community relations for the layer-coverage diversification in the local community being discovered.   
Our  experimental evaluation conducted on three real-world multilayer networks has shown   the significance of our approach.

\section{Multilayer Local Community Detection}
\label{sec:LCD}

\subsection{The \myalgo method}

We refer to the   multilayer network model described in~\cite{Kivela+14}. 
We are given    a set of layers  $\mathcal{L}$   
and    a set of entities (e.g., users)  $\V$.  We denote with    
$G_{\mathcal{L}} = (V_{\mathcal{L}}, E_{\mathcal{L}}, \V, \mathcal{L})$  the multilayer graph such that  $V_{\mathcal{L}}$ is a set of pairs  $v \in \V, L \in \mathcal{L}$,  and  $E_{\mathcal{L}} \subseteq  V_{\mathcal{L}} \times V_{\mathcal{L}}$ is the set of  undirected edges.         
Each entity of $V$ appears in at least one layer, but not necessarily in all layers.  
Moreover, in the following we will consider the specific case   for which nodes connected through different layers  the same entity in $\V$, i.e., $G_{\mathcal{L}}$ is a multiplex graph. 

Local community detection approaches generally implement some strategy that at each step  considers a node from one of three sets, namely: the community under construction (initialized with the seed node), the ``shell'' of   nodes that are neighbors of nodes in the community but  do not belong to the community, and the unexplored portion of the network.    
A key aspect is hence how to select the \textit{best} node in the shell to add to the community to be identified.   
Most algorithms, which are designed to deal with   monoplex  graphs, try to maximize a function in terms  of  the \textit{internal} edges, i.e., edges that involve nodes in the community, and to minimize a function in terms  of the  \textit{external} edges, i.e., edges to nodes outside the community. By accounting for both types of edges,   nodes that are candidates to be added to the community being constructed are penalized  in proportion to the amount of links to nodes external to the community~\cite{Clauset05}. 
Moreover,   as first analyzed in~\cite{ChenZG09}, considering  the internal-to-external \textit{connection density} ratio (rather than the absolute amount of internal and external links to the community)  allows for alleviating the issue of inserting many weakly-linked nodes (i.e.,  \textit{outliers}) into the local community being discovered. 
In this work we follow the above general approach  and extend it to identify local communities over a  multilayer  network.
  
Given  $G_{\mathcal{L}} = (V_{\mathcal{L}}, E_{\mathcal{L}}, \V, \mathcal{L})$ and a seed node $v_0$,  we denote with  $C \subseteq \V$   the node set   corresponding to the local community being discovered  around node $v_0$; moreover, when the context is clear, we might also use $C$ to refer to the local community subgraph.   
We denote with  $S = \{v \in  \V  \setminus C \ | \ \exists ((u,L_i),(v,L_j)) \in E_{\mathcal{L}} \ \wedge \ u \in C\}$  the \textit{shell} set of nodes outside $C$, 
and with 
$B = \{ u \in C  \  |  \ \exists  ((u,L_i),(v,L_j)) \in E_{\mathcal{L}} \ \wedge \ v \in S\}$ the \textit{boundary}  set of nodes in $C$. 

Our proposed  method, named {\em {\bf M}ulti{\bf L}ayer {\bf L}ocal {\bf C}ommunity {\bf D}etection} (\myalgo),    takes as input the multilayer graph $G_{\mathcal{L}}$ and a seed node $v_0$, and computes the local community $C$ associated to  $v_0$ by performing an iterative search that seeks to maximize the value of \textit{similarity-based local community function} for $C$ ($LC(C)$),  which is obtained as the ratio of an \textit{internal  community  relation} $LC^{int}(C)$ to an \textit{external community relation} $LC^{ext}(C)$. We shall  formally define these later in Section~\ref{sec:funcsim}.  
 
Algorithm \myalgo works as follows. Initially, the boundary set $B$ and the community  $C$  are initialized with the starting seed, while the shell set  $S$ is initialized with the neighborhood set of $v_0$ considering all the layers in  $\mathcal{L}$. 
Afterwards, the algorithm computes the initial value of $LC(C)$ and  starts   expanding the node set in  $C$:  
it evaluates all the nodes $v$ belonging to the current shell set $S$, then selects the vertex $v^{*}$ that maximizes the value of  $LC(C)$.  
The algorithm checks if  \textit{(i)} $v^{*}$ actually increases the quality of $C$ (i.e., $LC(C \cup \{v^{*}\})>LC(C)$) and \textit{(ii)} $v^{*}$ helps to strength the internal connectivity of the community (i.e., $LC^{int}(C \cup \{v^*\}) > LC^{int}(C) $). 
If both conditions are satisfied,   node $v^{*}$ is added to $C$ and the shell set is updated accordingly, otherwise   node $v^{*}$ is removed from $S$ as it cannot  lead to an increase in the value of $LC(C)$. In any case, the boundary set $B$ and $LC(C)$ are updated. The algorithm terminates when no further improvement in $LC(C)$ is possible.

\subsection{Similarity-based local community function}
\label{sec:funcsim}

To account for the multiplicity of layers,  we define the multilayer local community function $LC(\cdot)$ based on  a notion of similarity between  nodes.  In this regard, two major issues are how to choose the analytical form of the similarity function, and how to deal with the different, layer-specific connections that any  two nodes might have in the multilayer graph.  
We address   the first issue in an unsupervised fashion, by resorting to 
any similarity measure that can express the topological affinity of two nodes in a graph. 
Concerning the second issue, one straightforward solution is to determine the similarity between any two nodes focusing on each layer at a time. The above points are formally captured by the following definitions. 
%
We denote with $E^C$ the set of edges between nodes that belong to $C$ and with  $E_i^C$ the subset of  $E^C$    corresponding to edges in a given layer $L_i$.  Analogously, $E^B$ refers to the set of edges between nodes in $B$ and nodes in $S$, and  $E_i^B$ to its subset corresponding to $L_i$.  

Given a community $C$, we define the \textit{similarity-based local community function}  $LC(C)$ as the ratio between the \textit{internal  community  relation} and \textit{external community relation}, respectively  defined  as:
\begin{equation}\label{eq:Def2_Lin}
LC^{int}(C)=\frac{1}{|C|}\sum_{v \in C} \sum_{L_i \in \mathcal{L}} \sum_{\substack{(u,v) \in E_i^C  \ \wedge \ u \in C}} sim_i(u,v) 
\end{equation}
\begin{equation}\label{eq:Def2_Lex}
LC^{ext}(C) = \frac{1}{|B|} \sum_{v \in B}  \sum_{L_i \in \mathcal{L}} \sum_{\substack{(u,v) \in E_i^B  \ \wedge \ u \in S}} sim_i(u,v) 
\end{equation}

In the above equations, function $sim_i(u,v)$  computes   the similarity between any two nodes $u,v$  contextually to layer $L_i$. In this work, we define it in terms of  Jaccard coefficient, i.e.,
%
$sim_i(u,v)  = \frac{|N_i(u)  \cap N_i(v)|}{|N_i(u)  \cup N_i(v)|}$,  
%
where   $N_i(u)$   
 denotes the set of neighbors of node $u$ in     layer $L_i$.

\subsection{Layer-coverage diversification bias}
When discovering a  multilayer local community centered on  a seed  node, the iterative search process in \myalgo that  seeks to maximize the similarity-based local community measure, explores  the different layers of the network. This implies that  the various layers might contribute very differently from each other in terms of edges constituting the local community structure. 
In many cases, it can be desirable  to control the degree of heterogeneity of relations (i.e., layers)   inside the local community being discovered. 

In this regard, we  identify two main approaches:
\begin{itemize}
\item \textbf{Diversification-oriented approach.} 
 This approach relies on the assumption that a local community is better defined by increasing as much as possible the number of edges belonging to different layers.  
More specifically, we might want to  obtain a local community  characterized by high diversification in terms of presence of layers and variability of edges coming from different layers.  
\item \textbf{Balance-oriented approach.} Conversely to the previous case,  the aim is to produce a local community that shows a certain \textit{balance} in the presence of  layers, i.e.,   low variability of edges over the different layers.
This approach relies on the assumption that a local community  might be well suited to real cases when it is uniformly distributed among the different edge types  taken into account. 
\end{itemize}

  Following the above observations, here we    propose a methodology to incorporate a parametric control
of the layer-coverage diversification in the local community being discovered.  
To this purpose, we introduce a \textit{bias factor} $\beta$ in \myalgo which impacts on the node similarity  measure   according to the following logic:

\begin{equation}
\beta=
\begin{cases}
    (0, 1], &  \textit{diversification-oriented bias}\\
    0, & \textit{no bias}\\
    [-1,0), & \textit{balance-oriented bias}
\end{cases}
\end{equation}

\noindent
Positive values of $\beta$ push the community expansion process towards a diversi\-fication-oriented approach, and, conversely, negative $\beta$ lead to different levels of balance-oriented scheme. Note that   the \textit{no bias} case corresponds to handling the node similarity ``as is''. 
Note also that, by assuming values in a continuous range, at each iteration \myalgo is enabled to make a decision by accounting for a wider spectrum of degrees of layer-coverage diversification. 

Given  a node $v \in B$ and a node $u \in S$, for any $L_i \in \mathcal{L}$, 
we define the $\beta$-biased  similarity $sim_{\beta, i}(u,v)$  as follows:

\begin{eqnarray}
sim_{\beta, i}(u,v) = \frac{2sim_i(u,v)}{1+e^{-bf}},\\
bf=\beta[f(C \cup \{u\})-f(C)]
\end{eqnarray}

\noindent 
where $bf$ is a \textit{diversification factor} and $f(C)$ is a  function that measures the current diversification between the different layers in the community $C$; in the following, we assume it is defined as the standard deviation of the number of edges for each layer in the community. 
The difference $f(C \cup \{u\})-f(C)$ is positive when the insertion of node $u$ into  the community increases the coverage over a subset of layers, thus  diversifying  the presence of layers in the local community. Consequently, when $\beta$ is positive, the diversification effect is desired, i.e., there is a boost in the value of $sim_{\beta, i}$ (and vice versa for negative values of $\beta$).
Note that $\beta$ introduces a bias on the similarity between two nodes only when evaluating the inclusion of a shell node into a community $C$, i.e., when calculating $LC^{ext}(C)$. 

\section{Experimental Evaluation}
\label{sec:results}

We used three   multilayer network datasets, namely  
\textit{Airlines} (417 nodes corresponding to airport locations, 3588 edges, 37 layers corresponding to airline companies)~\cite{Cardillo}, \textit{AUCS} (61 employees as nodes, 620 edges, 5 acquaintance relations as layers)~\cite{Magnanibook}, and \textit{RealityMining} (88 users as nodes, 355 edges, 3 media types employed to communicate as layers)~\cite{KimL15}.  All network graphs are undirected, and inter-layer links are regarded as coupling edges.

\begin{table}[t!]
\caption{Mean and standard deviation size of  communities by varying $\beta$ (with step of 0.1).}
\centering
\scalebox{0.8}{
\begin{tabular}{|l|l||c|c|c|c|c|c|c|c|c|c|c|} 
\hline 
dataset & & -1.0 & -0.9 & -0.8 & -0.7 & -0.6 & -0.5 & -0.4 & -0.3 & -0.2 & -0.1 & \cellcolor{blue!15}0.0 \\ \hline
\multirow{ 2}{*}{\textit{Airlines}} & \textit{mean} & 5.73 & 5.91 & 6.20 & 6.47 & 6.74 & 7.06 & 7.57 & 8.10 & 9.13 & 10.33 & \cellcolor{blue!15}11.33 \\
& \textit{sd} & 4.68 & 4.97 & 5.45 & 5.83 & 6.39 & 6.81 & 7.63 & 8.62 & 10.58 & 12.80 & \cellcolor{blue!15}14.78  \\ \hline
\multirow{ 2}{*}{\textit{AUCS}} &  \textit{mean} & 6.38 & 6.59 & 6.64 & 6.75 & 6.84 & 6.85 & 6.92 & 7.13 & 7.16 & 7.77 & \cellcolor{blue!15}7.90  \\
 & \textit{sd} & 1.48 & 1.51 & 1.59 & 1.69 & 1.85 & 1.85 & 1.87 & 2.15 & 2.18 & 2.40 & \cellcolor{blue!15}2.74 \\ \hline
\textit{Reality-} & \textit{mean} & 3.21 & 3.24 & 3.25 & 3.25 & 3.32 & 3.32 & 3.34 & 3.34 & 3.34 & 3.37 & \cellcolor{blue!15}3.37 \\
\textit{Mining} & \textit{sd} & 1.61 & 1.64 & 1.66 & 1.66 & 1.73 & 1.73 & 1.74 & 1.74 & 1.74 & 1.77 & \cellcolor{blue!15}1.77 \\
\hline
\end{tabular} 
}
\\
\vspace{2mm}
\scalebox{0.8}{
\begin{tabular}{|l|l||c|c|c|c|c|c|c|c|c|c|} 
\hline 
dataset & &  0.1 & 0.2 & 0.3 & 0.4 & 0.5 & 0.6 & 0.7 & 0.8 & 0.9 & 1.0 \\ \hline
\multirow{ 2}{*}{\textit{Airlines}} & \textit{mean} &  9.80 & 9.02 & 8.82 & 8.37 & 8.20 & 7.93 & 7.53 & 7.26 & 7.06 & 7.06 \\
& \textit{sd} &  12.10 & 10.61 & 10.07 & 9.39 & 9.15 & 8.67 & 7.82 & 7.46 & 7.35 & 7.27 \\ \hline
\multirow{ 2}{*}{\textit{AUCS}} &  \textit{mean} &  8.77 & 8.92 & 8.92 & 8.89 & 8.89 & 8.89 & 8.87 & 8.85 & 8.85 & 8.85 \\
 & \textit{sd}  & 3.16 & 3.33 & 3.33 & 3.27 & 3.27 & 3.27 & 3.26 & 3.23 & 3.23 & 3.23 \\ \hline
\textit{Reality-} & \textit{mean}  & 3.38 & 3.39 & 3.39 & 3.39 & 3.36 & 3.36 & 3.32 & 3.18 & 3.17 & 3.17 \\
\textit{Mining} & \textit{sd}  & 1.78 & 1.78 & 1.78 & 1.78 & 1.74 & 1.74 & 1.71 & 1.60 & 1.59 & 1.59 \\
\hline
\end{tabular}
}
\label{tab:size}
\end{table}

\textbf{Size and structural characteristics of local communities.\ } 
We first analyzed the size of  the local communities extracted by \myalgo for each node. 
Table~\ref{tab:size} reports on the mean and standard deviation of the size of the  local communities by varying   of $\beta$. 
As regards the \textit{no bias} solution (i.e, $\beta=0.0$), largest  local communities correspond to  \textit{Airlines} (mean 11.33 $\pm$ 14.78), while medium size communities (7.90 $\pm$ 2.74) are  found for \textit{AUCS} and relatively small   communities (3.37 $\pm$ 1.77) for \textit{RealityMining}.
The impact of $\beta$ on the community size is roughly proportional to the number of layers, i.e., high on \textit{Airlines}, medium on \textit{AUCS} and low on \textit{RealityMining}. For \textit{Airlines} and \textit{AUCS}, smallest communities are obtained with the solution corresponding to $\beta=-1.0$, thus suggesting that the discovery process becomes more xenophobic (i.e., less inclusive) while shifting towards a balance-oriented scheme. 
Moreover, on  \textit{Airlines}, the mean size follows a roughly normal distribution, with most inclusive solution (i.e., largest size) corresponding to the unbiased one. A near normal distribution (centered on $0.2 \leq \beta \leq 0.4$) is also  observed for \textit{RealityMining}, while mean size values linearly increase with $\beta$ for \textit{AUCS}.

To understand the effect of $\beta$ on the structure of the local communities,  
we analyzed the distributions of per-layer mean \textit{average path length} and mean \textit{clustering coefficient} of the identified communities (results not shown). One major remark is  that  on the networks with a small number of layers, the two types of distributions tend to follow an increasing trend for balance-oriented bias (i.e., negative $\beta$), which becomes  roughly constant  for the diversification-oriented bias (i.e., positive $\beta$). On \textit{Airlines}, variability happens to be much higher for some layers, which in the case of mean average path length   ranges between 0.1 and 0.5 (as shown by a rapidly  decreasing trend for negative $\beta$, followed by a   peak for $\beta=0.2$,   then again a decreasing trend). 
 


\begin{figure}[t!]
\centering
\subfigure[\textit{Airlines}]{\includegraphics[width=0.32\columnwidth]{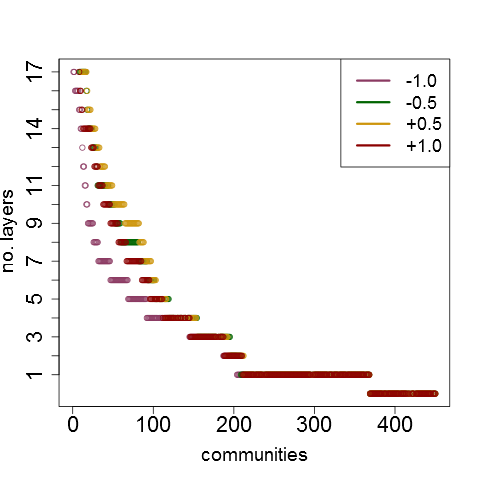}}
\subfigure[\textit{AUCS}]{\includegraphics[width=0.32\columnwidth]{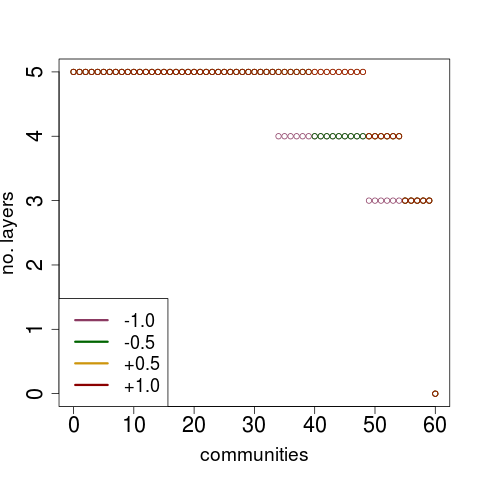}}
\subfigure[\textit{RealityMining}]{\includegraphics[width=0.32\columnwidth]{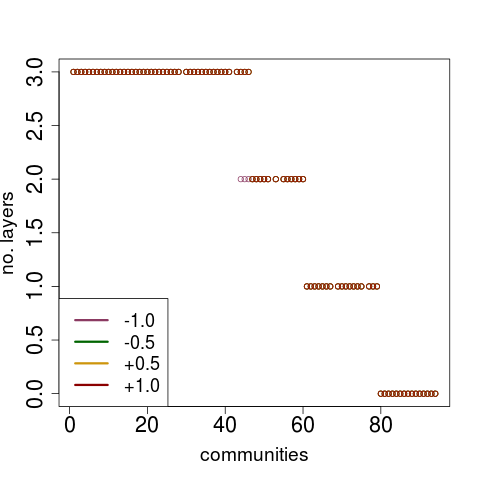}}
\caption{Distribution of number of layers over communities  by varying  $\beta$.  Communities are sorted by decreasing number of layers.} 
\label{fig:betalayers}
\end{figure}

\textbf{Distribution of layers over communities.\ } 
We also studied how the bias factor impacts on  the distribution of number of layers over communities, as shown in Figure~\ref{fig:betalayers}.  
This analysis confirmed that using positive values of $\beta$ produces  
local communities that lay  on a higher number of layers. This outcome can be easily explained since  positive values of $\beta$ favor the inclusion of nodes into the community which increase  layer-coverage diversification, thus enabling the exploration of  further layers also in an advanced phase of the discovering process. Conversely, negative values of $\beta$ are supposed to yield  a roughly uniform distribution   of the layers which are covered by  the community, thus preventing the discovery process from including nodes coming from unexplored layers once the local community is already characterized by a certain subset of layers. 


As regards the effects of the bias factor on the layer-coverage diversification, 
we analyzed the standard deviation of the per-layer number of edges by varying $\beta$  (results not shown, due to space limits of this paper). 
As expected, standard deviation values are roughly proportional to the setting of the bias factor for all datasets.
Considering the local communities obtained with negative $\beta$, the layers on which they lay are characterized by a similar presence (in terms of number of edges) in the induced community subgraph.
Conversely, for the local communities obtained using positive $\beta$, the induced community subgraph may be characterized by a small subset of layers, while other layers may be present with a smaller number of relations. 

\begin{figure}[t!]
\centering
\subfigure[\textit{Airlines}]{\includegraphics[width=0.45\columnwidth]{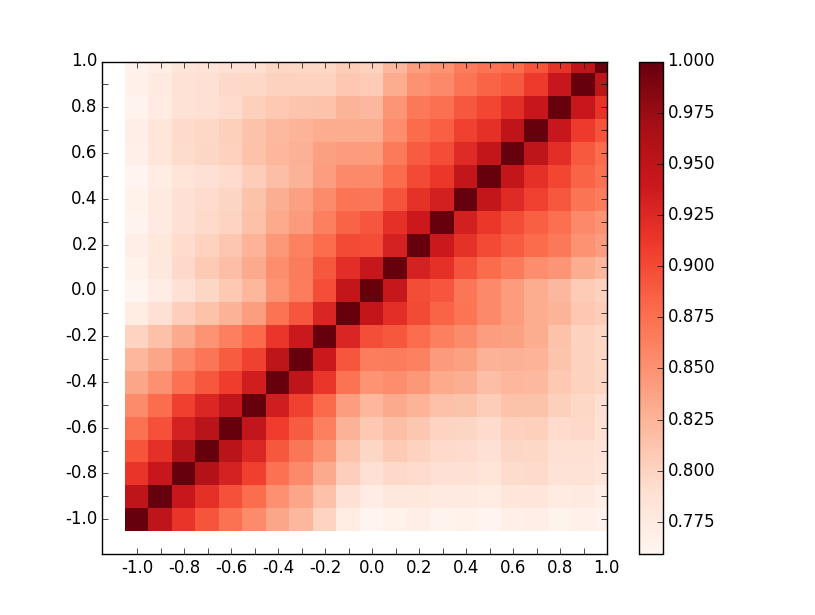}}
\subfigure[\textit{AUCS}]{\includegraphics[width=0.45\columnwidth]{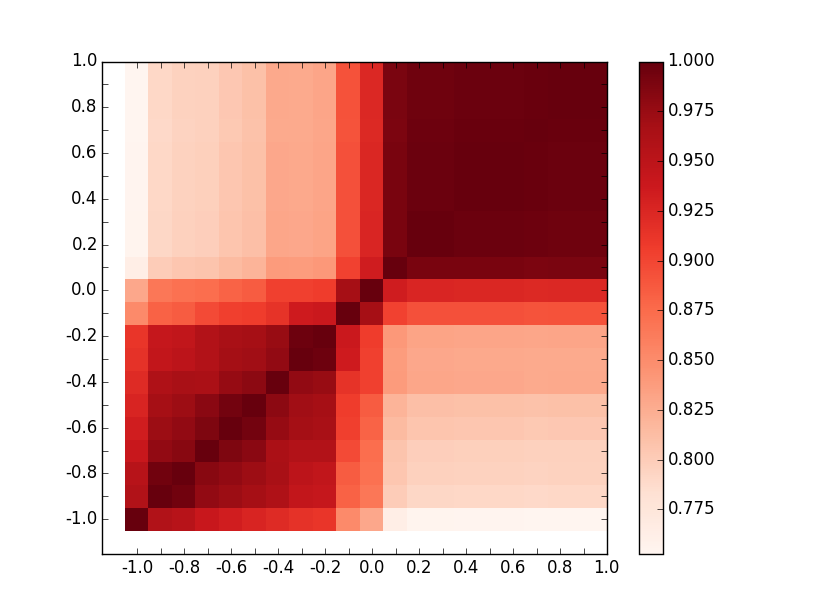}}
\caption{Average Jaccard similarity between solutions obtained by varying     $\beta$.} 
\label{fig:betalayers_jac}
\end{figure}

\textbf{Similarity between communities.\ } 
The smooth  effect due to  the diversification-oriented bias is confirmed when analyzing the similarity between the discovered  local communities. Figure~\ref{fig:betalayers_jac} shows the average Jaccard similarity between solutions obtained by varying  $\beta$ (i.e., in terms of nodes included in each local community). Jaccard similarities vary in the range $[0.75,1.0]$ for \textit{AUCS} and \textit{Airlines}, and in the range $[0.9,1.0]$ for \textit{RealityMining} (results not shown). 
For datasets with a lower number of layers (i.e., \textit{AUCS} and \textit{RealityMining}), there is a strong separation between the solutions obtained for $\beta>0$ and the ones obtained with $\beta<0$.
On \textit{AUCS}, the local communities obtained using a diversification-oriented bias show Jaccard similarities close to $1$, while there is more variability among the solutions obtained with the balance-oriented bias. Effects of the bias factor are lower on \textit{RealityMining}, with generally high Jaccard similarities. 
On \textit{Airlines},   the effects of the bias factor are still present but smoother, with gradual similarity variations in the range $[0.75,1.0]$.

\section{Conclusion}
\label{sec:conclusion}

We addressed the novel problem of local community detection in multilayer networks, providing a greedy heuristic that iteratively attempts to maximize  the internal-to-external connection density ratio by accounting for layer-specific topological information.  
Our method is also able to control the layer-coverage diversification in the local community being discovered, by means of  a  bias factor  embedded in   the similarity-based local community function.  
Evaluation was conducted on real-world multilayer networks.  
As future work,  we plan  to study  alternative objective functions for the ML-LCD problem. It would also be interesting to enrich the evaluation part  based on data with ground-truth information.   
We also envisage a number of application problems  for which   ML-LCD methods can profitably be used, such as  friendship prediction, targeted influence propagation, and more in general, mining in incomplete networks.

\end{document}